\documentstyle [12pt,epsfig]{article} 
\makeatletter

\newcommand{\CA}{{\rm C}$_{A}$ }
\newcommand{\CB}{{\rm C}$_{B}$ }
\newcommand{\CAP}{{\rm C}$_{A'}$ }
\newcommand{\CBP}{{\rm C}$_{B'}$ }

\@addtoreset{equation}{section}
\makeatother

\newcommand{\ba}{\begin{eqnarray}}
\newcommand{\ea}{\end{eqnarray}}

\newcommand{\MA}{$\mu_{\rm A}$ }
\newcommand{\MB}{$\mu_{\rm B}$ }
\newcommand{\MAP}{$\mu_{\rm A'}$ }
 
\begin{document}
\newcommand{\BS}{\bigskip}
\newcommand{\SECTION}[1]{\BS{\large\section{\bf #1}}}
\newcommand{\SUBSECTION}[1]{\BS{\large\subsection{\bf #1}}}
\newcommand{\SUBSUBSECTION}[1]{\BS{\large\subsubsection{\bf #1}}}

\begin{titlepage}
\begin{center}
\vspace*{2cm}
{\large \bf Muon decays in the Earth's atmosphere, time dilatation 
 and relativity of simultaneity}  
\vspace*{1.5cm}
\end{center}
\begin{center}
{\bf J.H.Field }
\end{center}
\begin{center}
{ 
D\'{e}partement de Physique Nucl\'{e}aire et Corpusculaire
 Universit\'{e} de Gen\`{e}ve . 24, quai Ernest-Ansermet
 CH-1211 Gen\`{e}ve 4.
}
\newline
\newline
   E-mail: john.field@cern.ch
\end{center}
\vspace*{2cm}
\begin{abstract}
   Observation of the decay of muons produced in the Earth's atmosphere by
    cosmic ray interactions provides a graphic illustration of the
    counter-intuitive space-time predictions of special relativity theory.
    Muons at rest in the atmosphere, decaying simultaneously, are
    subject to a universal time-dilatation effect when viewed from a
    moving frame and so are also
    observed to decay simultaneously in all such frames, whereas the decays
   of muons with different proper frames show relativity of simultaneity
    when observed from different inertial frames.

\vspace*{1cm}
\end{abstract}
\end{titlepage}
 
\SECTION{\bf{Introduction}}
    The present paper is the seventh in a recent series devoted to space-time physics, written by the
   present author, and posted on the arXiv preprint server.
   In Ref.~\cite{JHFRSPB} the classical `Rockets-and-String'
    ~\cite {RS} and `Pole-and-Barn'~\cite{PB} paradoxes of special relativity were re-analysed taking
    into account the distinction between the real and apparent\footnote{i.e. as naively predicted by 
   the standard space-time Lorentz transformation} positions of uniformly moving objects. Different
    results were obtained from the usual text-book interpretations of these experiments and a new
    causality-violating paradox was pointed out. This paradox, as well as the related `backwards running
    clocks' one of Soni~\cite{Soni}, was resolved in Ref.~\cite{JHFLLT} where, in order to avoid these
     paradoxes as well as manifest
     breakdown of translational invariance, in some applications of the standard space-time Lorentz
    transformation, the use of a `local' Lorentz transformation. i.e. one where the transformed event in
    the moving frame lies at the coordinate origin in this frame, was advocated. When this is done the
    closely correlated `relativity of simultaneity' (RS) and `length contraction' (LC) effects of 
    conventional special relativity theory do not
    occur. The connection between these effects is explained in Ref.~\cite{JHFRSPB}.
      Ref.~\cite{JHFLLT} also contains a `mini review' of all experimental tests of special relativity
      where it is pointed out that, whereas time dilatation is well-confirmed experimentally,
     no experimental evidence exists for the RS and LC effects. In the later papers
     ~\cite{JHFUMC} and ~\cite{JHFCRCS} it is explained how the spurious RS and LC effects result from a
     misuse of the time symbols in the standard space-time Lorentz transformation. Ref.~\cite{JHFCRCS} presents
    the argument in a pedagogical manner, whereas Ref.~\cite{JHFUMC} contains a brief `executive summary' of it.
     In Ref.~\cite{JHFAS} it is shown that the absence of the conventional RS effect
     follows directly from the Reciprocity Postulate~\cite{JHFLT,BG}, that holds in both Galilean and
      special relativity, without invoking the Galilean or Lorentz transformations. As discussed in
      Ref.~\cite{JHFFFT} mis-use of the  standard space-time Lorentz transformation in 
     classical electrodynamics leads to incorrect relativistic derivations, in particular of the classical
    (pre-relativistic) Heaviside formula~\cite{Heaviside} for the fields of a uniformly moving charge,
     which is no longer valid in relativistic classical electrodynamics. In Ref.~\cite{JHFTETE},
     Einstein's classic
     train/embankment experiment~\cite{EinTETE}, used in many
   text books to demonstrate the RS effect before introducing the Lorentz transformation, is reanalysed 
     and a conclusion  --the absence of any RS effect-- contrary to that of Einstein, is found. However 
    a genuine `relativity of simultaneity' effect is predicted by special relativity in the case that
   observers in two trains of suitably chosen speeds. as well as an embankment observer are considered.
   The main aim of the present paper is to present a conceptually similar, but simpler, demonstration of
    this RS effect (quite distinct from the text-book one derived by incorrect use of the Lorentz
    transformation) by considering the decay of muons at rest in different inertial frames.
   \par In the following section the necessary formulae for the analysis of the muon decay thought
    experiment --essentially the prediction of a universal time dilatation effect-- are derived from
    first principles. Here there is considerable overlap with work presented in ~\cite{JHFUMC} 
    and ~\cite{JHFCRCS}. The analysis of the thought experiment presented in Section 3 shows both the
    absence of the spurious text-book RS effect (muons which decay simultaneously
     in a common proper frame, are observed to do so in all inertial frames) as well as a genuine
     relativistic RS effect when the observation of the decays of muons at rest in different
     inertial frames is considered.
\SECTION{\bf{Operational meaning of the space-time Lorentz transformation: Rates and spatial
separations of moving clocks}}
 The Lorentz transformation (LT) relates observations ($x$,$y$,$z$,$t$) of the coordinates of space-time events in one inertial frame S,
  to observations of the coordinates
 ($x'$,$y'$,$z'$,$t'$) of the same events in another inertial frame S'.
   As is conventional, the Cartesian spatial coordinate
   axes of the two frames are parallel, and the origin of the frame S' moves with constant speed, $v$,
    along the $x$-axis.
   In any actual experiment, times are recorded by clocks and positions specified by marks on fixed
    rulers (or their equivalent). Therefore, in order to relate the space-time coordinates appearing in the
   LT to actual physical measurements they must be identified with clock readings and length interval
   measurements. This can be done in two distinct ways depending on whether the experimenter observing the
   clocks and performing the length measurements is at rest in S or in S'. In the former case (only events
   with spatial coordinates along the $x$-axis are considered so that $y$ = $y'$ = $z$ = $z'$ = $0$) the
   appropriate LT is:
    \begin{eqnarray}
  x'& = & \gamma_v[x-v\tau] \\
 t'& = & \gamma_v[\tau-\frac{\beta_v x}{c}]
\end{eqnarray}
 and in the latter case is:
    \begin{eqnarray}
  x& = & \gamma_v[x'+c\tau'] \\
 t& = & \gamma_v[\tau'+\frac{\beta_v x'}{c}]
\end{eqnarray}
 In these equations $\beta_v \equiv v/c$, $\gamma_v  \equiv 1/ \sqrt{1-\beta_v^2}$ and $c$ is the speed of
 light in vacuum. In (2.1) and (2.2) the transformed events lie on the worldline of a clock, C', at rest in S',
   which is observed from S. The observed time in S registered by C'( which is in motion in this frame)
  is $t'$, while $\tau$ is the time registered by a clock, C, identical to C', but at rest in S. In contrast,
  in (3) and (4) the transformed events lie on the worldline of C, which is observed from S'. The time
    $t$ is that registered by C as observed from S' and $\tau'$ is the time registered by C' as observed
   in its own proper frame. Thus two experiments are possible involving one stationary and one moving
    clock, depending on whether the experimenter performing the space and time measurements is in the rest
   frame of one, or the other, of the two clocks. To describe both of these experiments, the four different
    time symbols, $\tau$, $\tau'$, $t$ and $t'$, with different operational meanings, are required.
   \par In order to derive predictions, without introducing any specific spatial coordinate 
    system, it is convenient to introduce invariant interval relations~\cite{HP1906,Mink} that may be derived
   directly from (2.1) and (2.2) or (2.3) and (2.4):       
  \begin{equation}
      c^2  (\Delta t')^2 -  (\Delta x')^2
      = c^2  (\Delta \tau)^2 -  (\Delta x)^2 
   \end{equation}    
  \begin{equation}
      c^2  (\Delta t)^2 -  (\Delta x)^2
      = c^2  (\Delta \tau')^2 -  (\Delta x')^2 
   \end{equation} 
   (2.5) connects different events on the worldline of C', (2.6) different events
    on the worldline of C.  Since C' is at rest in S', $\Delta x' = 0$, while the equation of motion of
     C' in S is $\Delta x = v \Delta \tau$. Substituting these values in (2.5) and using the definition
   of $\gamma_v$, gives:
     \begin{equation}
     \Delta \tau = \gamma_v \Delta t'
  \end{equation}  
  Similarly, since C is at rest in S,  $\Delta x = 0$, and the equation of motion of C in S' is
  $\Delta x' = -v \Delta \tau'$. Hence (2.6) yields the relation:
     \begin{equation}
     \Delta \tau' = \gamma_v \Delta t
  \end{equation}  
    (2.7) and (2.8) are expressions of the relativistic Time Dilatation (TD) effect in the two
    `reciprocal' experiments that may be performed using the clocks C and C'. They show that, according,
    to the LT, `moving clocks run slow' in a universal manner (no spatial coordinates appear in (2.7) and (2.8)).
     In fact: 
   \begin{equation}
  \frac{{\rm rate~of~moving~clock}}{{\rm rate~of~stationary~clock}} =\frac{\Delta t'}{\Delta \tau}
     = \frac{\Delta t}{\Delta \tau'} =\frac{1}{\gamma_v}
   \end{equation}
   \par To discuss measurements of the spatial separations of moving clocks, at least two clocks,
    (say, \CAP and \CBP, at rest in S') must be considered. It is assumed that they lie along the $x'$-axis
     separated by the distance $L'$. It will be convenient to introduce two further identical clocks \CA and \CB
 at rest in S at such a separation that when the $x$-coordinates of \CA and \CAP coincide, so do those of
    \CB and \CBP. Further suppose that all the clocks are first stopped (i.e. no longer register time) and
    are set to zero. They are then all restarted at the instant that \CA and \CAP as well as \CB and \CBP
     have the same $x$-coordinates. Such a procedure has been introduced by Mansouri and Sexl~\cite{MS} and
    termed `system external clock sychronisation'. In this case, the time intervals in (2.7) and (2.8) may be 
    replaced by registered clock times: $\Delta \tau \rightarrow \tau$ etc. Also the space-like invariant
   interval relation connecting arbitary points on the wordlines of \CAP and \CBP may be written as:
       \begin{eqnarray}     
  (\Delta s)^2  & \equiv & (L')^2 -c^2 [t'({\rm C}_{B'})-t'({\rm C}_{A'})]^2 \nonumber \\
     &  = & [x({\rm C}_{B'}) - x({\rm C}_{A'})]^2-c^2 [\tau({\rm C}_{B})-\tau({\rm C}_{A})]^2 
    \end{eqnarray}    
   The spatial separation of \CAP and \CBP in the frame S, $L$, is defined as the value of
  $x({\rm C}_{B'}) - x({\rm C}_{A'})$ at some particular instant in S. For this
   $\tau({\rm C}_{B}) = \tau({\rm C}_{A})$, so that (2.10) may be written:
 \begin{equation}     
  (\Delta s)^2   \equiv  (L')^2 -c^2 [t'({\rm C}_{B'})-t'({\rm C}_{A'})]^2 =  L^2 
  \end{equation}
  
     After the `external synchronisation' procedure (2.7) may be written:
    $\tau({\rm C}_{A}) = \gamma_v t'({\rm C}_{A'})$ and  $\tau({\rm C}_{B}) = \gamma_v t'({\rm C}_{B'})$
    for the clocks \CAP and \CBP respectively. Then, if $\tau({\rm C}_{B}) = \tau({\rm C}_{A})$, 
     necessarily $ t'({\rm C}_ {B'}) =  t'({\rm C}_{A'})$ and (2.11) simplifies to:
     \begin{equation}
       \Delta s \equiv L' = L
    \end{equation}
     The measured spatial separation of the clocks is therefore a Lorentz invariant quantity~\cite{JHFPS}
      that is the same in all inertial frames. Thus there is  no `relativistic length contraction' effect.
      Also, since $\tau({\rm C}_{B}) = \tau({\rm C}_{A})$  requires that also
       $t'({\rm C}_ {B'}) =   t'({\rm C}_{A'})$ there is here no `relativity of simultaneity' effect either.
      How these correlated, spurious, effects arise from misinterpretation of meaning of the time symbols
      in the LT is explained elsewhere~\cite{JHFLLT,JHFUMC,JHFCRCS,JHFAS}. A genuine relativity of simultaneity
       effect arising in observations of muon decays, occuring in different inertial frames, is, however, described
     in the following section. The clocks A', A and B, discussed in an abstract fashion above, are physically
       realised by
      introducing the muons \MAP, \MA and \MB. Each muon constitutes an independent clock whose proper time, $T$,
     is signalled in an arbitary inertial frame by observation of the muon decay event: $\mu \rightarrow e \nu
      \bar{\nu}$.

 \SECTION{\bf{Muons are clocks that demonstrate time dilatation and relativity of simultaneity}}
    Muon decays constitute an excellent laboratory for testing the predictions of special relativity.
    For example, the TD effect of Eqn(2.7) was experimentally
    confirmed at the per mille level of relative precision in the ultrarelativisic domain 
    ($\gamma_v \simeq 30$) by observation of the decay of muons in the
    storage ring of the last CERN muon $g-2$ experiment~\cite{NatureTD}.
    In the present paper, it is shown that thought experiments involving muons provide a graphic illustration
    of the predicted space-time behaviour, in special relativity, of clocks in different inertial frames.
    \par  Unlike most other unstable particles, muons are particularly suitable for
       precise tests of the TD effect because of the ease of their production from pion
       decay and long mean lifetime of 2.2 $\mu$s. The former yields high events statistics and the latter
       the possiblity of precise time interval measurements using accurate clocks in the
       laboratory frame~\cite{NatureTD}.
       \par The thought experiment developed in the present paper is an elaboration of the well-known
        demonstration that the very presence of cosmic muons at the Earth's surface is, by itself,
     sufficient to demonstrate the existence of the TD effect~\cite{FL,TW,TL,CC}. Muons are produced predominantly
      by the weak decay of charged pions $\pi^{\pm} \rightarrow \mu^{\pm} \nu$. The velocity of the muon, 
      $v_{\mu}$, depends upon that of the parent muon, $v_{\pi}$, and
    the centre-of-mass decay angle, $\theta^{\star}$.
      If the pion has the same velocity, $v_{\mu}^{\star} = c(m_{\pi}^2-m_{\mu}^2)/(m_{\pi}^2+m_{\mu}^2)$,
      as the muon in the pion rest frame, (corresponding to a pion momentum of 49.5 MeV/c)
      and $\cos \theta^{\star} = -1$, the muon is produced at rest in the laboratory system. 
      The maximum muon decay energy $E_{\mu}^{max}$ correponds to  $\cos \theta^{\star} = 1$ and is given, 
     in terms of the parent pion energy $E_{\pi}$, and the pion velocity $v_{\pi} = c \beta_{\pi}$, by the
      relation:
      \begin{equation}
     E_{\mu}^{max} = E_{\pi} \frac{[m_{\pi}^2(1+\beta_{\pi})+m_{\mu}^2(1-\beta_{\pi})]}{2 m_{\pi}^2}
       \end{equation} 
   
      For ultra-relativistic parent pions with $\beta_{\pi} \simeq 1$, $ E_{\mu}^{max} \simeq  E_{\pi}$.
     \par Due to the thickness of the Earth's atmosphere, the majority of interactions of primary
    cosmic protons, that produce the parent pions of cosmic muons, occur at high altitude, $\simeq$ 20 km above the Earth's surface. A muon with 
      speed close to that of light then takes at least $\simeq$ 700 $\mu s$ to reach the surface of the Earth.
     This may be compared with the muon mean lifetime of 2.2 $\mu s$. Without the TD effect, only
      a fraction $\exp[-700/2.2] \simeq 10^{-138}$ of the muons produced at altitude would reach
      the Earth's surface. However a 10 GeV muon, which has $\gamma_v \simeq 94$, has a 3.5 $\%$
       probability to
      reach the  Earth's surface, before decaying, when the TD effect is taken into account. 
      \par In the thought experiment considered here it is assumed that two muons \MA and \MAP
       are produced simultaneously at the same point A (see Fig.1a) by decay of pions from a primary cosmic
    ray interaction with the nucleus of a gas atom of the atmosphere. The muon \MA is produced at rest in
    the atmosphere (inertial frame S) while \MAP is produced with velocity $v = c \beta_v = \sqrt{3}/2$, so that
     $\gamma_v = 2$. It happens that both muons decay after time $T$ in their proper frames.
      Because of the TD effect, the muon \MAP will then be seen by an observer at rest in
     the atmosphere to decay after time $\tau = \gamma_v T = 2T$ at a point B at a distance 
     $L = 2Tv = 2.28$km from A. It is also supposed that at the same time, $\tau = 0$,  that \MA and \MAP
     are created another muon, \MB, (also with proper decay lifetime $T$) is created at rest in the
      atmosphere at the 
    point B, by decay of pion from another primary cosmic ray interaction. Since \MA and \MB are at rest in 
     the atmosphere and have no TD effect, they will decay
    simultaneously at $\tau = T$ (Fig.1b) in the frame S. At this instant the muon \MAP is still undecayed and is
    at the point M, midway between A and B, When \MAP decays (Fig.1c)  \MA and \MB no longer
    exist, however the centres of mass of their, by now distant, decay products $e$, $\nu$ and $\bar{\nu}$ 
    still remain at the points A and B. 
    \par The sequence of events that would be seen by an observer in the rest frame, S', of \MAP
      is shown in Fig.2. In S', \MA and \MB move to
     the left with velocity $v$. The configuation at $\tau = \tau' = 0$ is shown in Fig.2a. At time 
     $\tau' = T = L/(2v)$ (Fig.2b) \MAP decays when it has the same $x'$ coordinate as the point M.
     The muons \MA and \MB are still undecayed. At time $\tau' = \gamma_v T = L/v$ (Fig.2c) \MA and \MB are
     observed to decay simultaneously. At this time A'(the centre of mass of the \MAP decay products)
      has the same $x$-coordinate as the point B in the
     atmosphere. Note that \MA and \MB are observed to decay
     simultaneously in S' as well as in S. In fact their decays will
     be observed to be simultaneous in {\it any} inertial frame. In
     this case, there is no RS effect as predicted by a conventional
     (but incorrect) text book application of the LT~\cite{JHFLLT,JHFUMC,JHFCRCS,JHFAS}.

\begin{figure}[htbp]
\begin{center}\hspace*{-0.5cm}\mbox{
\epsfysize15.0cm\epsffile{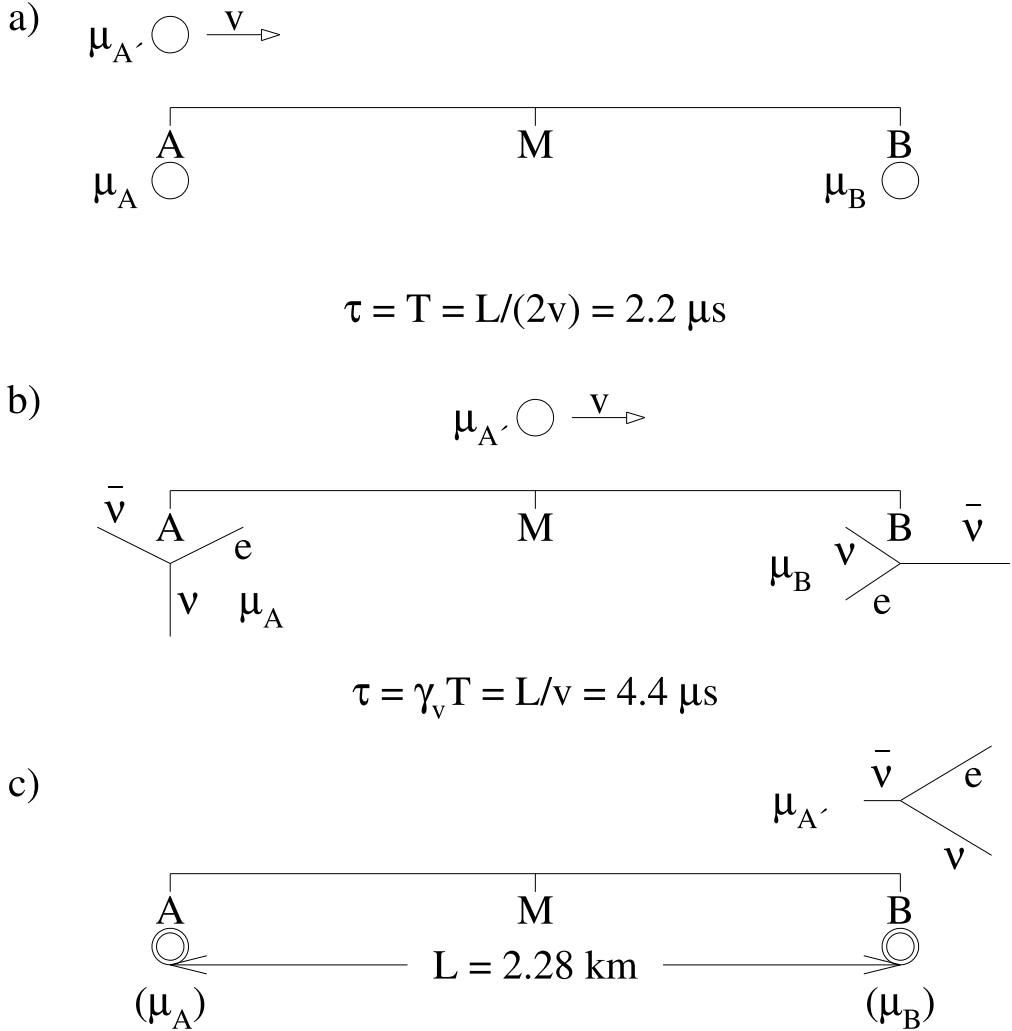}}
\caption{ {\em  The sequence of muon decay events as observed from the atmosphere (frame S).
  a) Muons \MAP, \MA and \MB are simultaneously created. Muon \MAP moves to the right with
 velocity $v = (\sqrt{3}/2)c$. b) At time $\tau = T$, muons \MA and \MB decay simultaneously.
  At this time \MAP is observed from S to be aligned with the mid-point, M, of A and B.
  c) At time $\tau = \gamma_v T$, muon \MAP is observed to decay. At this time it is at B,
 the centre of mass of the decay products of \MB. For clarity, the muons are shown displaced vertically.}}  
\label{fig-fig1}
\end{center}
\end{figure}

\begin{figure}[htbp]
\begin{center}\hspace*{-0.5cm}\mbox{
\epsfysize15.0cm\epsffile{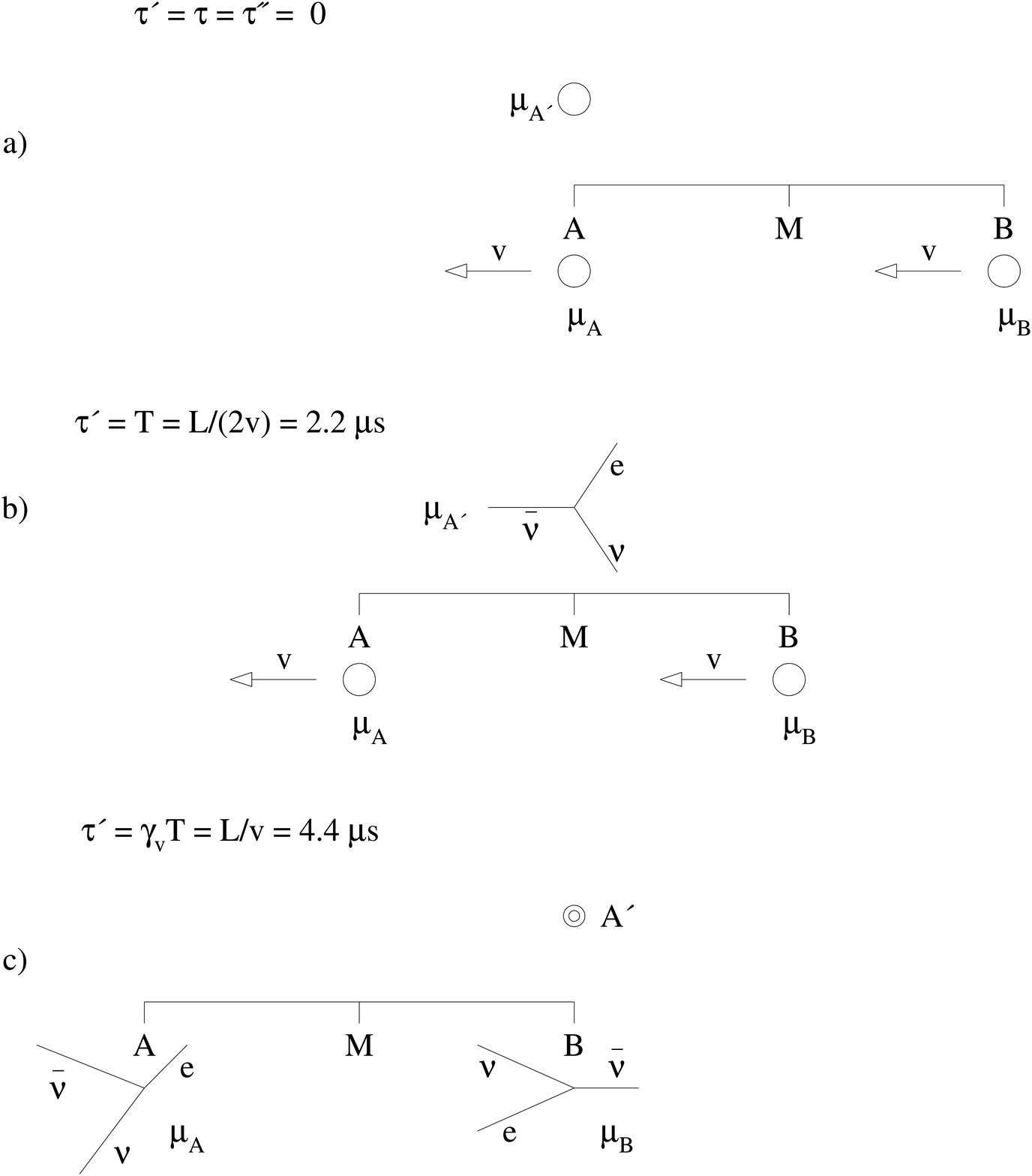}}
\caption{{\em The sequence of muon decay events as observed in the proper frame (S') of \newline \MAP.
  a) Muons \MAP, \MA and \MB are simultaneously created. Muons \MA and \MB  move to the left with
 velocity $v = (\sqrt{3}/2)c$. b) At time $\tau' = T$, muon \MAP decays.
  At this time it is aligned with the mid-point, M, of A and B.
  c) At time $\tau' = \gamma_v T$ muon \MA and \MB decay simultaneously. At this time B,
 is aligned with the point A', the centre of mass of the decay products of \MAP.
 For clarity, the muons are shown displaced vertically.}}
\label{fig-fig2}
\end{center}
\end{figure}

\begin{figure}[htbp]
\begin{center}\hspace*{-0.5cm}\mbox{
\epsfysize15.0cm\epsffile{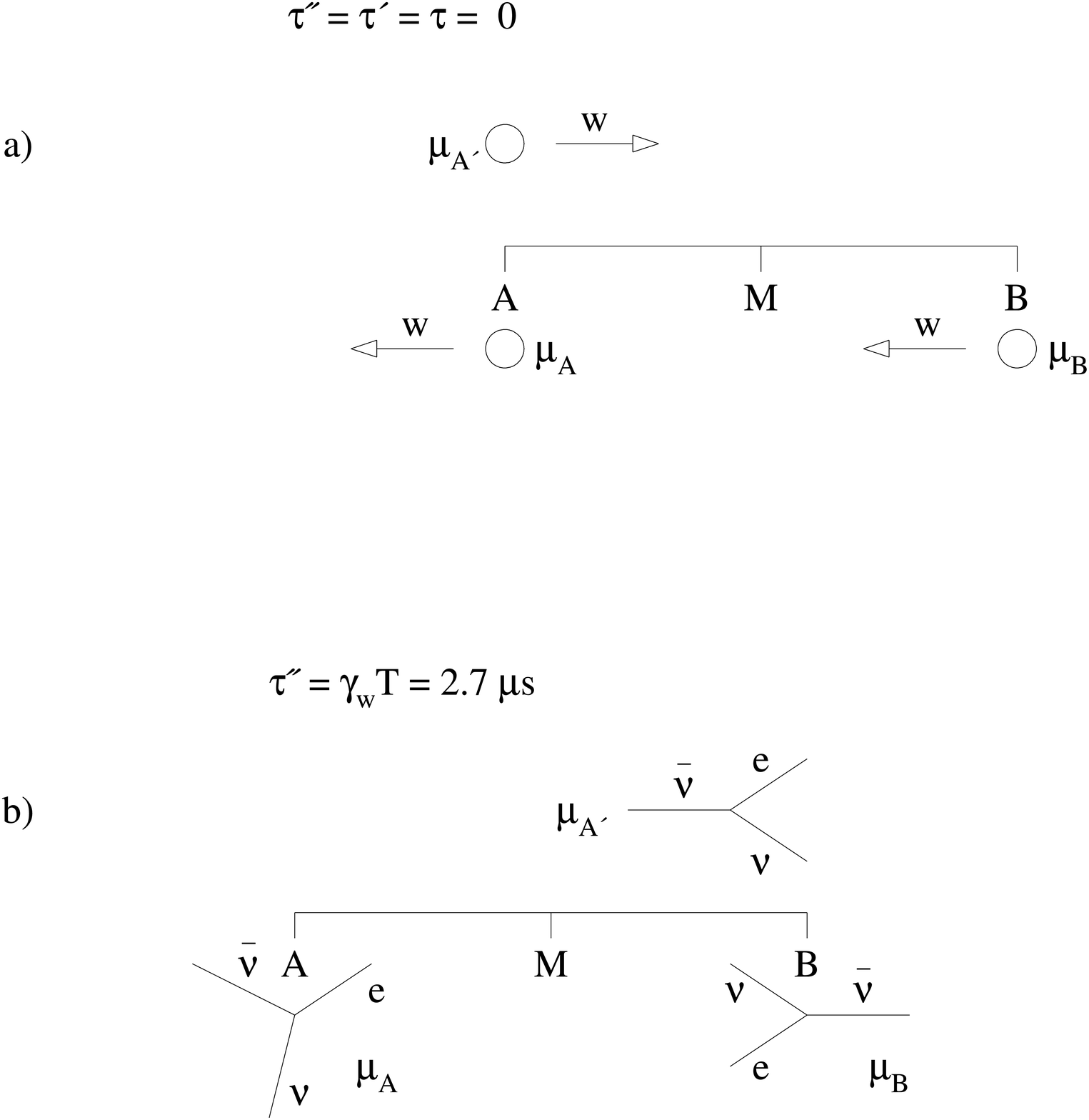}}
\caption{{\em  The sequence of muon decay events as observed from the frame (S'') that
  moves parallel to the direction of motion of \MAP with the velocity $w = c^2(\gamma_v-1)/(v\gamma_v) =c/\sqrt{3}$.
 a) Muons \MAP, \MA and \MB are simultaneously created. Muon \MAP moves to the right with
  velocity $w$ and \MA and \MB  move to the left with
 velocity $w$. b) At time $\tau'' = \gamma_w T$, all three muons decay simultaneously.
    In this case, unlike for the reciprocal observations shown in Figs.1c and 2c, \MAP and \MB(or the
   centres of mass of their decay products) are not aligned at the time of the decay of either.
   For clarity, the muons are shown displaced vertically.}}
\label{fig-fig3}
\end{center}
\end{figure}
    
    \par Finally, in Fig.3, is shown the same sequence of muon decay events as they would be seen
      by an observer at rest in the frame S'' that is moving parallel to the direction of motion
    of \MAP with velocity $w = c^2(\gamma_v-1)/(v \gamma_v)$ relative to the atmosphere. In S'',
   \MAP moves with speed $w$ in one direction, while \MA and \MB move with speed $w$ in the
    opposite one (see Fig.3a). Since the TD effect is now the same for all three muons they will
    observed from S'' to decay simultaneously at the time $\tau'' = \gamma_w T = 0.613 L/v$ as shown in
    Fig.3b.
    \par The muon decay events thus exhibit a genuine `relativity of simultaneity' effect.
     The simultaneous decays of \MA and \MB in their proper frame S (fig.1b) are also seen to be
     simultaneous in the frames S' (Fig.2c) and S'' (Fig.3b). However, the decay times $\tau_D$,
      $\tau'_D$. and  $\tau''_D$ of \MAP show relativity of simultaneity relative to those of \MA and \MB:
      \begin{eqnarray}
       {\rm Frame~S~(Fig.1)~~~} \tau_D(\mu_{{\rm A'}}) & >  & \tau_D(\mu_{{\rm A}}) =  \tau_D(\mu_{{\rm B}})
        \nonumber \\
       {\rm Frame~S''~(Fig.3)~~} \tau''_D(\mu_{{\rm A'}}) & = & \tau''_D(\mu_{{\rm A}}) =  \tau''_D(\mu_{{\rm B}})           \nonumber \\ 
      {\rm Frame~S'~(Fig.2)~~} \tau'_D(\mu_{{\rm A'}}) & < & \tau'_D(\mu_{{\rm A}}) =  \tau'_D(\mu_{{\rm B}})
     \nonumber
       \end{eqnarray}
     A similar pattern of events is seen in a variant of Einstein's train/embankment thought experiment
     with trains moving at speeds $v$ and $w$ relative to the embankment~\cite{JHFTETE}. 

 \par Table 1 shows the decay times and relative spatial separations of the muons
  \MAP and \MB (or the centres of mass of their decay products) in the frames S (the atmosphere, or
   the proper frame of \MB),
   S' (the \MAP proper frame) and S'' (a frame moving with velocity $w$ relative to the atmosphere).
   The symbols $\tilde{\tau}$ and $\tilde{x}$ denote generic proper times or coordinates. e.g.
   $\tilde{\tau}$ stands for any of $\tau$, $\tau'$ and  $\tau''$. The differences of proper times
   of the decay events in different frames or the spatial separation of \MAP and \MB at the decay times
   of either, in these frames, are all of O($\beta^2$), and so vanish in the Galilean limit $c \rightarrow \infty$.
    The times and separations in this Galilean limit (the same in all frames) are shown in the last row of
   Table 1. Note that the limit of $w$ as $c \rightarrow \infty$ is $v/2$, so that the spatial separations
   $x''_D(\mu_{{\rm A'}})-x''_{{\rm B}}$ and $x''_D(\mu_{{\rm B}})-x''_{{\rm A'}}$ both vanish in this limit.
   \par Inspection of, and reflection upon, Figs.1-3 shows behaviour greatly at variance with intuition derived
    from Galilean space-time. The same muon decay events appear differently ordered in time and at different
   spatial separations depending on the frame of observation. These effects are the space-time analogues 
    of the distortions produced by linear perspective in visual perception. Just as the latter
    are unique for every observer, so the former are unique to every different inertial frame. 

  \begin{table}
\begin{center}
\begin{tabular}{|c||c c c c|} \hline
 Frame   &\multicolumn{1}{c|}{ $\tilde{\tau}_D(\mu_{{\rm A'}})$}
    &\multicolumn{1}{c|}{$\tilde{x}_D(\mu_{{\rm A'}})-\tilde{x}_{{\rm B}}$}
 &\multicolumn{1}{c|}{$\tilde{\tau}_D(\mu_{{\rm B}})$}
 &\multicolumn{1}{c|}{$\tilde{x}_D(\mu_{{\rm B}})-\tilde{x}_{{\rm A'}}$} \\ \hline \hline
     &\multicolumn{4}{c|}{\em Special Relativity} \\ \cline{2-5}
  S & $\gamma_v T$  & 0 & $T$    & $v(\gamma_v-1)T$   \\
 S' &  $T$   &  $-v(\gamma_v-1)T$   &  $\gamma_v T$  & 0 \\
 S'' &  $\gamma_w T$  & $-(v \gamma_v-2 w \gamma_w)T$  & $\gamma_w T$ 
 &  $(v \gamma_v-2 w \gamma_w)T$ \\ \cline{2-5}
      &\multicolumn{4}{c|}{\em Galilean Relativity} \\ \cline{2-5}
 S,S',S'' & $T$ & 0 & $T$ & 0 \\   
 \hline 
\end{tabular}
\caption[]{{\em Decay times and spatial separations of muons \MAP and \MB, or the centres of mass of
  their decay products, in the frames S, S' and S''. $\tilde{\tau}$ and $\tilde{x}$ are generic proper 
  times and spatial coordinates. The last row shows the predictions of Galilean
   relativity, that are the same in all frames.}}      
\end{center}
\end{table}

\par {\bf Added Note}
 \par  A revised and re-titled version of the present paper is available~\cite{JHFMUDECR}. Calculational and
  conceptual errors in the original paper~\cite{JHFMUDECV2} are explained in this note.
   \par The calculation of the invariance of length intervals based on the invariant interval relation
    (2.10) is flawed since the times $t'({\rm C}_{A'})$, $t'({\rm C}_{B'})$ do not correspond to those of
    synchronised clocks in the frame S'. Placing ${\rm C}_{A'}$ at the origin of S' and ${\rm C}_{B'}$
     at $x' = L'$ the correct LT equations for clocks which are synchronised in S' are:
    \begin{eqnarray}
  x'({\rm C}_{A'}) & = & \gamma_v[x({\rm C}_{A'})-v\tau({\rm C}_{A})] = 0 \\
 t'({\rm C}_{A'}) & = & \gamma_v[\tau({\rm C}_{A}-\frac{\beta_v x({\rm C}_{A'})}{c}] \\
  x'({\rm C}_{B'})-L' & = & \gamma_v[x({\rm C}_{B'})-L-v\tau({\rm C}_{B})] = 0 \\
 t'({\rm C}_{B'}) & = & \gamma_v[\tau({\rm C}_{B})-\frac{\beta_v (x({\rm C}_{B'})-L)}{c}]
\end{eqnarray}
   These equations show that  $t'({\rm C}_{A'}) = t'({\rm C}_{B'}) = \tau({\rm C}_{A}) = \tau({\rm C}_{B}) = 0$
    when $x({\rm C}_{A'})= x({\rm C}_{B'})-L = 0$, so that all four clocks are synchronised at this instant.
    The correct invariant interval relation is then not (2.10) but:
         \begin{eqnarray}     
  (\Delta s)^2  & \equiv & [x'({\rm C}_{B'}) - x'({\rm C}_{A'})]^2 -c^2 [t'({\rm C}_{B'})-t'({\rm C}_{A'})]^2 \nonumber \\
     &  = & [x({\rm C}_{B'}) - x({\rm C}_{A'})]^2-c^2 [\tau({\rm C}_{B})-\tau({\rm C}_{A})]^2 \nonumber \\
     &    & +2 [x({\rm C}_{B'}) - x({\rm C}_{A'})](\gamma_v L'-L) - 2 v \gamma_v[\tau({\rm C}_{B})-\tau({\rm C}_{A})]
   \nonumber \\  
     &    & +L^2-2 \gamma_v L L'+ (L')^2
    \end{eqnarray}
   Setting $ \tau({\rm C}_{B})=\tau({\rm C}_{A})$ and hence, from (2.7), $t'({\rm C}_{B'})=t'({\rm C}_{A'})$ in (3.6)
   gives not the relation $L = L'$ but instead the trivial identity:
   \begin{equation}
      \Delta s = L' = L'
   \end{equation}
    In fact, the equality $L = L'$ is already implicit in in (3.4) which uses the same spatial coordinate systems in S and S'
     as (3.2). Since $L = x({\rm C}_{B'},\tau({\rm C}_{B})= 0)$ independently of the value of $v$, Eqn(3.4) is valid for 
    all values of $v$. In particular it holds when $v = 0$, $\gamma_v = 1$ and $x \rightarrow x'$, in which case it is written:
   \begin{equation}
     x'({\rm C}_{B'})-L' =  x'({\rm C}_{B'})-L 
   \end{equation}
  so that
  \begin{equation}
    L' = L 
   \end{equation}
    \par A major conceptual error occurs in the interpretation of Figs.2 and 3. It is assumed that these
      represent observations in the frames S' and S'' of the events defined in the frame S in Fig.1. That is, 
     observations of the same events in different frames in the same space-time experiment. If this were indeed 
      the case, then use of the relative velocity transformation formula, Eqn(2.14) of Ref.~\cite{JHFMUDECR}
      shows that the speed of  \MA and \MB should  be $v\gamma_v$ in Fig.2 and  $w\gamma_w$ in Fig. 3. It then
     follows that
     the claimed `relativity of simultaneity' effect is annuled and the stated mismatch of spatio-temporal
     coincidence events (for example, \MAP is aligned with B in the frame S, and M in the frame S', when it decays)
     does not occur. Indeed, as previously correctly pointed out~\cite{Langevin,Mermin}, the same spatio-temporal
     event must occur in all frames from which it may be observed. 
    \par What are actually shown in Fig.2 and Fig.3 are the configurations of {\it physically independent experiments}
      in these frames. The configuration in Fig. 2 is that of an experiment which is reciprocal to that of Fig. 1
       with TD effect given By Eqn(2.8) (clocks in S seen to run slower than clocks in S' in both frames)  not
      that of the primary experiment described by (2.7) in which clocks in S' are seen to run slower than clocks in S
      in both frames. The configuration of Fig.3 is that of yet another space-time experiment related to that of
      Fig.1, by a boost with velocity $w$ in the positive $x$-direction, in which the clocks associated with all the muons are observed to run at the same rate,
      and to decay simultaneously, in all three frames. 
       \par In conclusion, when the experiment is correctly analysed, no `relativity of simultaneity'  or
        `length contraction´ effects occur
          and the same spatio-temporal coincidence events, at the epochs of the muon decays, are observed in all frames,
          as they must be.

\pagebreak

\end{document}